\documentclass[10pt,a4paper,twoside]{article}
\usepackage{epsfig}
\usepackage{baltlat5}
\usepackage{wrapfig}
\begin{document}
\ \
\vspace{0.5mm}

\setcounter{page}{1}
\vspace{5mm}

\titlehead{Baltic Astronomy, vol.\ts 14, XXX--XXX, 2005.}

\titleb{The peculiar sdB NGC\,6121-V46: A low-mass double
  degenerate ellipsoidal variable in a globular cluster}


\begin{authorl}
\authorb{S.~J.~O'Toole}{1}
\authorb{R.~Napiwotzki}{2}
\authorb{U.~Heber}{1}
\authorb{H.~Drechsel}{1}
\authorb{S.~Frandsen}{3}
\authorb{F.~Grundahl}{3}
\authorb{H.~Bruntt}{4}
\end{authorl}

\begin{addressl}
\addressb{1}{Dr. Remeis-Sternwarte, Astronomisches Institut der Universit\"at Erlangen-N\"urnberg, Sternwartstr. 7, Bamberg 96049, Germany}

\addressb{2}{Centre for Astrophysics Research, Univ. of
  Hertfordshire, Hatfield AL10 9AB, UK}

\addressb{3}{Department of Physics and Astronomy, Univ. of
  Aarhus, Ny Munkegade, DK-8000 Aarhus C, Denmark}
\addressb{4}{Niels Bohr Institute, Juliane Maries Vej 30, DK-2100
  Copenhagen \O, Denmark}
\end{addressl}

%

\submitb{Received 2005 August 1}

\begin{abstract}
The variable sdB known as V46 in the globular cluster M4 has remained
enigmatic since its discovery almost 10 years ago. We present here
radial velocity measurements obtained from medium-resolution VLT/FORS2
spectra that show variations at twice the period of the luminosity
changes. This implies that the system is an ellipsoidal
variable. Unlike the other sdB binaries of this nature, the
fundamental parameters of this star we derive suggest that it lies
below the Zero Age Extreme Horizontal Branch. From the cluster
distance and the gravity we determine the mass of V46 to be
$\sim$0.19M$_\odot$. This is too low to sustain core helium
burning. From the mass function we derive a lower limit for the
companion of only 0.26M$_\odot$. We discuss the star's origin in the
context of close binary evolution in the field and globular clusters. 
\end{abstract}




\begin{keywords}
stars: variable: general -- stars: individual: NGC6121-V46
\end{keywords}

\resthead{The peculiar sdB NGC\,6121-V46}{S.~J.~O'Toole et al.}

\sectionb{1}{History and Motivation}

It is now known that while the mass distribution of DA white dwarfs
peaks at $\sim$0.6M$_\odot$, there is also a subset of objects with masses
$\le$0.46M$_\odot$ (Bergeron et al. 1992). Their low mass implies that
they cannot ignite helium in their cores, and must have lost most of
their envelope mass before reaching the tip of the red giant
branch. The implication of this is that the stars must be in close
binary systems, although recent results suggest that fewer than 50\%
show radial velocity variations (Napiwotzki et al. 2005).

There are two kinds of known binaries containing helium-core white
dwarfs (HeWDs): double degenerate systems, typically containing two low-mass
white dwarfs (e.g. Nelemans \& Tout 2005) and millisecond pulsar (MSP)
systems. In the former case the lowest mass HeWD companion known has
0.31M$_\odot$ (Marsh et al.\ 1995), while in the latter the masses
are $\le$0.2M$_\odot$ (e.g. Callanan et al.\ 1998). 


Recent discoveries suggest a link between HeWDs and a small subset of
sdB stars. %
As part of a program to
measure radial velocities of sdBs, Heber et al. (2003) discovered that
the apparently normal sdB HD\,188112 lies below the Zero-Age Extreme
Horizontal Branch (ZAEHB). The  star's mass, determined from its
trigonometric parallax, $T_{\mathrm{eff}}$ and $\log g$ to be only
0.24M$_\odot$, is inconsistent with the lowest mass for the common
envelope ejection model of Han et al. (2003). HD\,188112 is a radial
velocity variable with a minimum companion mass of 0.73M$_\odot$,
suggesting the primary is the progenitor of a helium core white dwarf,
while the unseen companion is most likely a C/O white
dwarf. Comparison with the evolutionary tracks of Driebe et al. (1998)
supports this conclusion. 

Since then, Liebert et al. (2004) have discovered another object lying
below the ZAEHB. Comparison of the parameters of
SDSS\,J123410.37-022802.9 (hereafter SDSS\,J1234) with the evolution
tracks of Althaus et al. (2001) shows that this star has a mass in the range
0.18-0.19M$_\odot$. Note that the mass is dependent on the evolution models
adopted. A higher mass is found ($\sim$0.23M$_\odot$) for SDSS\,J1234
when it is compared to the Driebe et al. tracks. 

Kaluzny et al. (1997) discovered intensity variations with
amplitude $\sim$1.9\% and period 1.045 h in the star now known as
V46 in NGC\,6121 (=\,M4). The system was a puzzle from the beginning,
with colours consistent with those of a cataclysmic variable, but no
emission lines present in its spectrum. Mochejska et al. (2002) found
that the spectrum is consistent with an sdB star. With the discovery of
pulsations in sdB stars with periods of 0.75-2 hours (Green et
al. 2003), it seemed possible that V46 was a member of this class. We
therefore set out to clarify the object's nature.

\vskip1mm


\begin{wrapfigure}[14]{r}[0pt]{60mm}
\vskip-15mm
\vbox{
\vskip2mm
\centerline{\psfig{figure=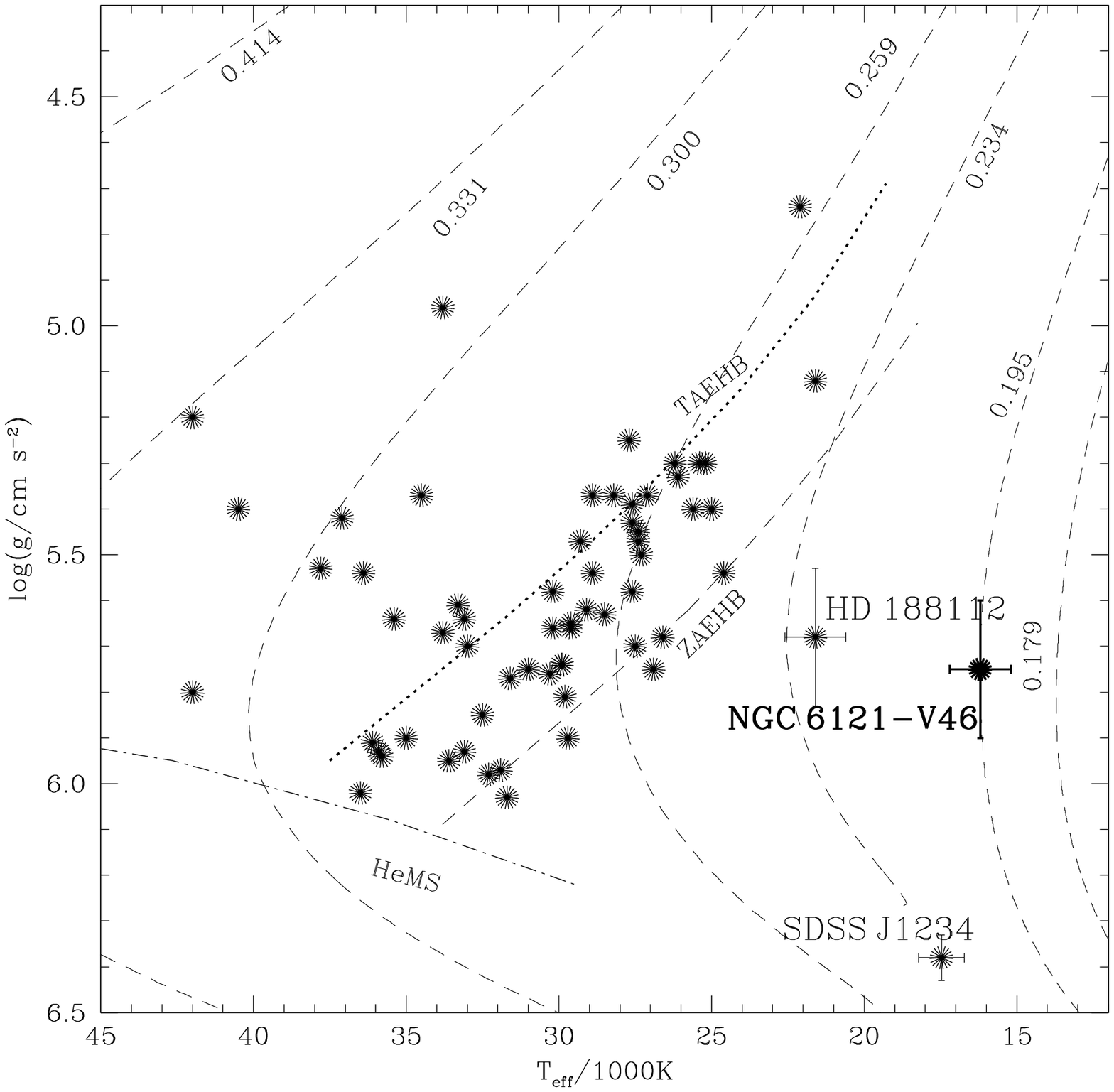,width=50mm,angle=0,clip=}}
\vskip-15mm
\captionc{1}{Position of V46 in the $T_{\mathrm{eff}}-\log g$ diagram.}
}
\end{wrapfigure}
\vskip2mm

\sectionb{2}{Is it an EHB star?}

A low resolution spectrum of V46 taken with the MMT was kindly
provided by Janus Kaluzny. By means of Balmer line profile fitting we
find $T_{\mathrm{eff}}=16197\pm546$\,K and $\log g=5.75\pm0.108$; the
star shows no helium lines. We used Detlev Koester's DA white dwarf
grid (Finley et al. 1997) to derive these parameters.
When we examine the position of V46 in the
$T_{\mathrm{eff}}-\log g$ plane (Figure 1), we find that it, like
HD\,188112 and SDSS\,J1234, lies below the ZAEHB. This means the star
has not evolved via the EHB and strongly suggests that it too is the
progenitor of a HeWD (similar to HD\,188112).

If we consider the distance to the cluster M4 (2.2 kpc -- Rosenberg et
al. 2000) and the surface gravity of V46, we find that it should have
a mass of $\sim$0.19M$_\odot$. In Figure 1 we show the post-RGB
evolutionary tracks of Driebe et al. (1998) which suggest that V46
should have a mass of $\sim$0.195M$_\odot$. This is very good
agreement considering the uncertainties of the model physics, and is
strong evidence that the star is a member of the cluster and not a
foreground object. Using the VLT we have determined the nature of
this system by measuring the radial velocity of V46.




\sectionb{3}{The Nature of NGC\,6121-V46}

We obtained 81 spectra with resolution $\sim$2100 using VLT/FORS2
with the 1400V grating in service mode. The only line visible in the
spectra is H$\beta$, and the resulting S/N
per 4-minute exposure was 10-15. We measured the radial
velocity of H$\beta$ by cross correlation with a model spectrum at the
temperature and gravity derived above. The velocity curve is shown at
the top of Figure 2, while the power spectrum is shown at the
bottom. It was calculated by fitting the curve with a range of periods
using $\chi^2$ minimisation. Based on this method, we measure the period and
ephemeris to be HJD($T_0$)$=2453134.496797+0.087159\times E$, while the
system velocity and semi-amplitude of the system are
$\gamma=31.3\pm1.6$\,km/s and $K=211.6\pm2.3$\,km/s,
respectively. Using this period and semi-amplitude we derive the mass
function of the system to be $f(m)=0.0855\pm$\,0.0028M$_\odot$. Taking the
mass of the sdB to be 0.195M$_\odot$, the minimum mass of the
secondary is 0.26M$_\odot$. 
We note that the period of the radial velocity variation
is exactly twice that of the intensity variation, indicating that
V46 is an ellipsoidal variable, similar to the previously known
systems KPD 0422+5421 (Koen et al. 1998) and KPD 1930+2752
(Bill\'eres et al. 2000; Maxted et al. 2000). 

\begin{wrapfigure}[14]{r}[0pt]{82mm}
\vskip-15mm
\vbox{
\vskip2mm
\centerline{\psfig{figure=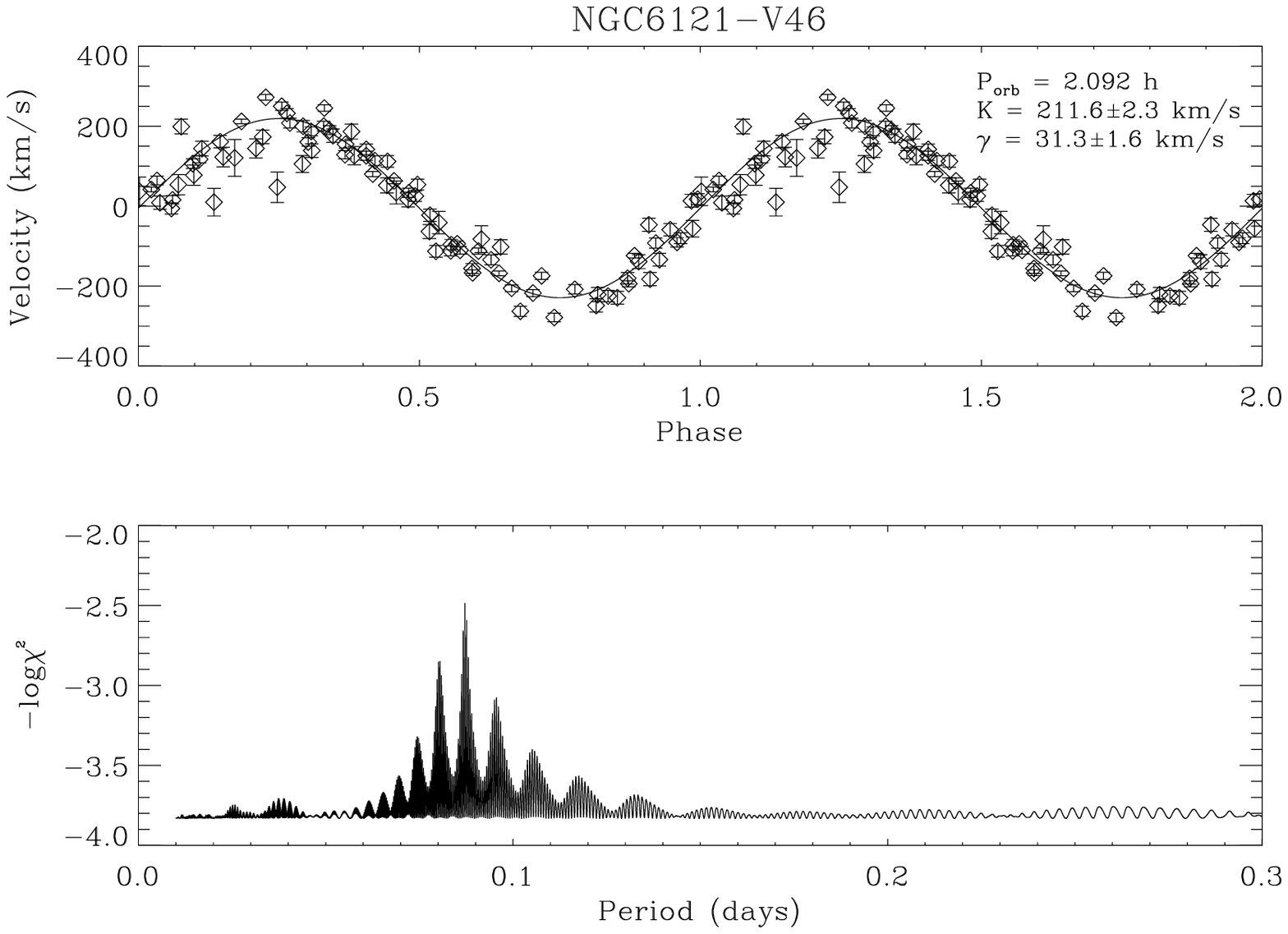,width=70truemm,angle=0,clip=}}
\vskip-17mm
\captionc{2}{Radial velocity curve and corresponding power spectrum of
  V46.}
}
\end{wrapfigure}
\vskip2mm


\vskip3mm

\sectionb{4}{Light Curves}

As well as a velocity curve, we have also obtained BVR light
curves. These observations were made over April-June 2001 to search
for oscillations in red giants in M4 (Bruntt 2003). They were obtained
at the Danish 1.54m at La Silla, Chile with DFOSC. We have attempted
to determine the inclination angle of the system using a light curve
analysis similar to that used for the sdB HS 2333+3927 (Heber et
al. 2004).  No eclipses have been detected. Due to the low amplitude
of the ellipsoidal variations the analysis is difficult. We kept the
$T_{\mathrm{eff}}$ and mass of the sdB fixed at 16\,200\,K and
0.2M$_\odot$, respectively, during the analysis, but could not
find a unique solution. Even fixing the mass ratio arbitrarily at,
e.g. $q=3.0$, does not help. The fits from all
solutions appear identical and have very similar $\chi^2$
values. There must be a lower limit for the inclination since at some
angle the deformation of the sdB will no longer be visible. To find
this limit, we first set $q=6.0$ and then decreased it in steps of
1.0. Again no unique solution could be found, meaning that
unfortunately the light curves cannot place any constraints on the
system. Nevertheless we can conclude that the companion must be a
white dwarf, because the variations are ellipsoidal in
nature and there is no sign of a reflection effect. The latter would
be expected if it were a low-mass main sequence star.

\vskip3mm

\sectionb{5}{Implications}

Despite not being able to constrain the inclination of the V46 system,
we can still discuss the implications of its existence for close
binary formation in globular clusters. Hansen et al. (2003) discussed
low-mass HeWDs in globular clusters from a
theoretical standpoint, and found that for the population of these
stars in the core of NGC 6397 the companions are most likely C/O white
dwarfs and not neutron stars. The likelihood of the companion of V46
being a neutron star is not high, since the inclination angle of the
system would have to be $\le26\degr$, and there has been no detection at
radio wavelengths of any neutron star in M4 other than PSR\,B1620-26
(Lyne et al. 1988). We suggest therefore that V46's companion is
either a C/O or HeWD. At the most probable
inclination of 52$\degr$ the mass of the companion is
$\sim$0.40M$_\odot$, making it an object with a helium core. We also
point out that V46 lies outside the central part of M4; more massive
objects are expected to settle towards the core of a globular cluster.



\goodbreak


\References

\refb
Althaus~L.~G., Serenelli ~A.~M., Benvenuto~O.~G. 2001, MNRAS, 323, 471

\refb
Bergeron~P., Saffer~R.~A., Liebert~J. 1992, ApJ, 394, 228

\refb
Bill\'eres~M., Fontaine~G., Brassard~P., et al. 2000, ApJ, 530, 441

\refb
Bruntt~H. 2003, PhD dissertation, University of Aarhus

\refb
Callanan~P.~J., Garnavich~P.~M., Koester~D. 1998, MNRAS, 298, 207

\refb
Driebe~T., Sch\"onberner~D., Bl\"ocker~T., Herwig~F. 1998, A\&A, 339, 123

\refb
Finley~D.~S., Koester~D., Basri~G. 1997, ApJ, 488, 375

\refb
Green~E.~M., Fontaine~G., Reed~M.~D., et al. 2003, ApJ, 583, L31

\refb
Han~Z., Podsiadlowski~P., Maxted~P.~F.~L., Marsh~T.~R. 2003, MNRAS,
341, 669

\refb
Hansen~B.~M.~S., Kalogera~V., Rasio~F.~A. 2003, ApJ, 586, 1364

\refb
Heber~U., Drechsel~H., {\O}stensen~R., et al. 2004, A\&A, 420, 251

\refb
Heber~U., Edelmann~H., Lisker~T., Napiwotzki~R. 2003, A\&A, 411, L477

\refb
Kaluzny~J., Thompson~I.~B., Krzeminski~W. 1997, AJ, 113, 2219

\refb
Koen~C., Orosz~J.~A., Wade~R.~A. 1998, MNRAS, 300, 695

\refb
Liebert~J., Bergeron~P., Eisenstein~D., et al. 2004, ApJ, 606, L147

\refb
Lyne~A.~G., Biggs~J.~D., Brinklow~A., McKenna~J., Ashworth~M. 1988,
Nature, 332, 45

\refb
Marsh~T.~R., Dhillon~V.~S., Duck~S.~R. 1995, MNRAS, 275, 828

\refb
Maxted~P.~F.~L., Marsh~T.~R., North~R.~C. 2000, MNRAS, 317, L41

\refb
Mochejska~B.~J., Kaluzny~J., Thompson~I., Pych~W. 2002, AJ, 124, 1486

\refb
Nelemans~G., Tout~C.~A. 2005, MNRAS, 356, 753


\refb
Rosenberg~A., Piotto~G., Saviane~I., Aparicio~A. 2000, A\&AS, 144, 5

\vskip10mm


\end{document}